\documentclass[english]{article}
\usepackage[T1]{fontenc}
\usepackage[latin9]{inputenc}
\usepackage[letterpaper]{geometry}
\usepackage{array}
\usepackage{textcomp}

\makeatletter

\providecommand{\tabularnewline}{\\}

\@ifundefined{definecolor}
 {\usepackage{color}}{}
\makeatother

\makeatother

\usepackage{babel}

\begin{document}

\title{Improved protocols of secure quantum communication using $W$ states}

\author{Chitra Shukla$^{1}$, Anindita Banerjee$^{2}$ and Anirban Pathak$^{1,3}$}

\maketitle
\begin{center}
$^{1}$Jaypee Institute of Information Technology, A-10, Sector-62,
Noida, UP-201307, India
\par\end{center}

\begin{center}
$^{2}$Department of Physics and Center for Astroparticle Physics
and Space Science, Bose Institute, Block EN, Sector V, Kolkata 700091,
India.
\par\end{center}

\begin{center}
$^{3}$RCPTM, Joint Laboratory of Optics of Palacky University and
Institute of Physics of Academy of Science of the Czech Republic,
Faculty of Science, Palacky University, 17. listopadu 12, 77146 Olomouc,
Czech Republic.
\par\end{center}
\begin{abstract}
Recently, Hwang \emph{et al}. {[}Eur. Phys. J. D. \textbf{61, }785
(2011){]} and Yuan \emph{et al}. {[}Int. J. Theo. Phys. \textbf{50,
}2403 (2011){]} have proposed two efficient protocols of secure quantum
communication using 3-qubit and 4-qubit symmetric $W$ state respectively.
These two dense coding based protocols are generalized and their efficiencies
are considerably improved. Simple bounds on the qubit efficiency of
deterministic secure quantum communication (DSQC) and quantum secure
direct communication (QSDC) protocols are obtained and it is shown
that dense coding is not essential for designing of maximally efficient
DSQC and QSDC protocols. This fact is used to design maximally efficient
protocols of DSQC and QSDC using 3-qubit and 4-qubit $W$ states.
\end{abstract}

\section{Introduction}

In 1984, Bennett and Brassard proposed a protocol \cite{bb84} for
quantum key distribution (QKD), which allows two remote legitimate
users (Alice and Bob) to establish an unconditionally secure key through
the transmission of qubits. Since then several protocols for different
cryptographic tasks have been proposed \cite{ekert,b92,review,Imoto,Hillery,ping-pong,vaidman-goldenberg,lm05}.
While most of the initial works on quantum cryptography \cite{bb84,ekert,b92}
were limited to QKD only. Eventually the idea got extended to direct
secure quantum communication \cite{review} where the legitimate users
can communicate directly without establishing any prior key. The protocols
of direct secure quantum communication are broadly divided into two
classes \cite{review}: A) Protocols of deterministic secure quantum
communication (DSQC), where the receiver can read out the secret message
encoded by the sender, only after the transmission of at least one
bit of additional classical information for each qubit. B) Protocols
of quantum secure direct communication (QSDC), which does not require
any such exchange of classical information. In 1999, Shimizu and Imoto
\cite{Imoto} provided the first protocol of DSQC using Bell states.
But this pioneering work did not draw much attention of the quantum
cryptography community in the context of direct secure quantum communication.
In 2002 Bostrom and Felbinger \cite{ping-pong} proposed a Bell state
based QSDC protocol, which is popularly known as ping-pong protocol
and in 2005 Lucamarini and Mancini \cite{lm05} proposed a QSDC protocol
(LM05 protocol) without using entangled states. These two protocols
have drawn considerable attention. In the ping-pong protocol \cite{ping-pong}
Bob prepares a Bell state (say $|\psi^{+}\rangle$), keeps one photon
as home photon and sends the other photon to Alice as travel photon.
After receiving the travel photon Alice encodes bit value $0$ ($1$)
by applying $I$ ($X)$ on the travel qubit and sends it back to Bob.
Bob does a Bell measurement on the final state. If he obtains $|\psi^{+}\rangle$
then Alice's encoded bit is $0$ and if he obtains $|\phi^{+}\rangle$
then Alice's encoded bit is $1$. Here one can easily recognize that
the full power of dense coding is not used. Alice could have used
$I,\, X,\, iY$ and $Z$ to encode $00,01,10$ and $11$ respectively
and that would have increased the efficiency of ping-pong protocol.
This is so because the same amount of communication would have successfully
carried two bits of classical information. This fact was first formally
included in a modified ping-pong protocol proposed by Cai and Li in
2004 \cite{cai and li PRA}. This simple idea of inclusion of dense
coding to increase the efficiency of a secure direct communication
protocol has considerably influenced the future development of QSDC
and DSQC protocols. To be precise, Deng \emph{et al}. \cite{Deng protocol}
modified Cai and Li's dense coding based two way protocol into a dense
coding based one way two step protocol of QSDC (DLL protocol), where
Alice prepares large number of Bell states, keeps the home photons
(one photon of each entangled pair) with herself and sends the travel
photons to Bob. Now Bob measures half of the photons received by him
randomly in $X$ basis or $Z$ basis and announces the outcome of
his measurements, basis used and the position of the photons. Then
Alice measures the corresponding photons using the same basis. This
detects eavesdropping. In absence of eavesdropping, Alice encodes
her secret message on part of the remaining home photons (a fraction
of the home photons is kept for checking of eavesdropping) and sends
all the home photons to Bob. Finally, a Bell measurement by Bob successfully
decodes the secret message of Alice. The information encoded states
have to be mutually orthogonal, otherwise Bob would not be able to
discriminate them with certainty, i.e. Bob would not be able to deterministically
decode the information encoded by Alice. Thus Alice can not send the
encoded orthogonal states directly through the quantum channel because
in that case Eve will also be able to decode the information without
being detected. So the information is sent in two steps. This logic
indicates that two steps are essential for secure direct communications.
Keeping this in mind, several authors proposed DLL type one way two
steps dense coding based QSDC protocols using different entangled
states. Now a question comes in our mind: Is there any other way in
which the secure direct communication protocol can be made one way
one step protocol. There exist a clever trick for this purpose. The
trick, which is known as {}``rearrangement of order of particles''
was first introduced by Deng and Long for QKD in 2003 \cite{rearrangement of particle order}.
In 2006 Zhu \emph{et al}. \cite{reordering1} explicitly provided
a DSQC protocol using rearrangement of order of particles. In a Zhu
\emph{et al}. \cite{reordering1} type of protocol that uses rearrangement
of order of particles, we are allowed to send the entire information
encoded quantum state in one step. This type of protocols are analogous
to Deng type of protocols, with only difference that after encoding
operation Alice changes the sequence of particles and inserts some
decoy photons (prepared in non-orthogonal states) for eavesdropping
checking and sends this modified sequence to Bob. After Bob confirms
that he has received all the photons, Alice announces the position
of decoy photons and then checks eavesdropping by measuring the decoy
photons. In absence of Eve, Alice discloses the actual order. Even
if Eve is present and measure all the particles, she obtains only
a random sequence of bits since the order of the particles are rearranged.
Thus in this type of protocols encoded states are sent in one step.

Here it would be apt to note that, all protocols of QKD, DSQC and
QSDC essentially involve splitting of information into 2 or more pieces.
Having each piece by itself should be non-revealing of encoded bit.
This splitting of information can be done in several ways. Here one
can easily observe that in Goldenberg-Vaidman (GV) \cite{vaidman-goldenberg}
ping-pong (PP) \cite{ping-pong}, Cai-Li (CL)\cite{cai and li PRA}
and Deng-Long-Liu (DLL) \cite{Deng protocol} protocols the information
is split into two quantum pieces but in BB-84 and rearrangement
of particle ordering based protocols it is divided into a quantum
piece and a classical piece. To be precise, in DLL, PP, CL and other
protocols of QSDC we first check that the first quantum piece of information
is delivered to the receiver without any eavesdropping. Only when
this is ensured then the encoding operation is done. Consequently
Eve can never have access to both pieces of information and an individual
piece is non revealing by itself. Similarly, in rearrangement of particle
ordering based protocols the encoded quantum states, which is sent
first, is the quantum piece and the sequence of particles, which is
sent at the end, is the classical piece. For a successful decoding
we need a simultaneous access to both the pieces. Bob has this required
simultaneous access but Eve does not have it as she can not withhold
the quantum piece and wait for the announcement of the classical piece.
This is the secret of secrecy in rearrangement of particle order based
protocols. This interesting and nice trick is correctly used in most
of the recent protocols of DSQC%
\footnote{Since in this type of protocols Alice needs to announce exact sequence
(classical information) so these are examples of DSQC.%
} \cite{the:high-capacity-wstate,the:C.-W.-Tsai}. But there exist
examples of recent proposals where the information splitting is not
done properly. For example, in Guo \emph{et al}. \cite{guo} and some
other proposals required rearrangement of particle ordering is not
done. This has made these protocols insecure. Interestingly, all the
recently proposed DSQC protocols \cite{the:high-capacity-wstate,the:C.-W.-Tsai,hwang-hwang-Tsai,guo}
also use dense coding operations for encoding of information. The
coupling between dense coding and efficient protocols of DSQC and
QSDC went so strong that people started thinking that it will not
be possible to design maximally efficient DSQC/QSDC using $W$ states
since maximal dense coding is not possible in them. Keeping this in
mind several authors have designed inefficient (non-maximally efficient)
protocols of DSQC and QSDC using $W$ states (\cite{the:high-capacity-wstate}
and references there in) and have considered their protocols as efficient.
Contrary to this belief, here we will show that dense coding is not
necessary for implementation of maximally efficient DSQC and QSDC
protocols. We will further show that it is possible to design maximally
efficient DSQC and QSDC protocols using $W$ state and without using
dense coding. Before we describe our final results it would be apt
to briefly review some of the recent developments on DSQC and QSDC
using symmetric $W$ states. We have done the same in the next section.
In Section \ref{sec:Existing-DSQC-and} we have also provided two
new schemes for dense coding of 4-qubit $W$ states. Further, it is
shown that one of these two new schemes or one of the two existing
schemes \cite{wang and yan,pati1} of dense coding can be directly
used to increase qubit efficiency of the existing DSQC/QSDC protocols.
But efficiency of such protocols will not be maximal as the dense
coding is not maximal. In Section \ref{sec:Generalized-protocol-of}
we have provided generalized protocol of DSQC which is in general
valid for $n$-qubit symmetric $W$ states but the explicit Tables
of encoding operations are provided for $n=3$ and $n=4$ only. It
is shown that the dense coding is not required for the implementation
of the protocol. The proposed protocol is further generalized to a
QSDC protocol and its relation to QKD is described. The security of
the proposed protocol is also described in this section. In Section
\ref{sec:Efficiency-analysis} upper bounds on the qubit efficiency
\cite{defn of qubit efficiency} of the DSQC and QSDC protocols are
obtained in general and it is shown that the proposed DSQC protocol
and its QSDC counterpart are maximally efficient. The efficiency is
also compared with the existing protocols that uses symmetric $W$
states. Finally, Section \ref{sec:Conclusions} is dedicated for conclusions.

\section{\label{sec:Existing-DSQC-and}Existing DSQC and QSDC protocols using
symmetric $W$ states.}

$W$ states are rigorously studied as an important resource for quantum
communication tasks since a decade (see \cite{hwang-hwang-Tsai,the:high-capacity-wstate,opt com-2010,wang,cao and song}
and references there in). Our purpose is not to discuss all the proposals
of quantum information processing using $W$ states rather our focus
is only on DSQC and QSDC protocols. A protocol of DSQC using 4-qubit
$W$ state was first proposed by Cao and Song in 2006 \cite{cao and song}%
\footnote{The authors had claimed it as a QSDC protocol but it is actually a
DSQC protocol since Bob needs the measurement outcomes of Alice's
measurement to decode the classical information encoded by Alice.%
}. The Cao and Song protocol can transmit one bit of classical information
using a 4-qubit $W$ state. Thus the Cao and Song protocol is not
really efficient. Similarly, in the subsequent years, several protocols
of DSQC using 3-qubit $W$ states were also proposed. These protocols
can only transmit one bit of classical information using 3-qubit $W$
state \cite{wang-Zhang-tang,dong 1}. In recent past a considerably
large number of DSQC and QSDC protocols were proposed using 3-qubit
$W$ state \cite{hwang-hwang-Tsai,opt com-2010,yuan-two-step} and
using 4-qubit $W$ state \cite{the:high-capacity-wstate,guo}. These
new proposals \cite{the:high-capacity-wstate,hwang-hwang-Tsai,opt com-2010,yuan-two-step}
are usually compared with the inefficient initial protocols \cite{cao and song,wang-Zhang-tang,dong 1}
and shown to be efficient. Recently proposed protocols are definitely
better than the initial protocols (e.g. Cao and Song protocol \cite{cao and song})
but are still inefficient. This can be understood quickly if we consider
the fact that so called high-capacity DSQC protocol of Yuan \emph{et
al}. \cite{the:high-capacity-wstate} and the very recently proposed
DSQC protocol of Guo \emph{et al.} \cite{guo} uses 4-qubit $W$ state
to transmit 2 bits of classical information while an exactly similar
protocol \cite{the:C.-W.-Tsai} proposed by Tsai \emph{et al}. can
transmit 4 bits of classical information using 4-qubit cluster state.
As these protocols use dense coding operations and maximal dense coding
is allowed in cluster state but not in $W$ state \cite{pati1} so
it seems natural that cluster state based protocol is more efficient.
But a careful look into the encoding operations of Yuan \emph{et al.}
protocol can easily reveal that they have not even used the full power
of dense coding. Two independent schemes for dense coding using 4-qubit
W states are reported by Pradhan \emph{et al}. \cite{pati1} and Wang
and Yan \cite{wang and yan}. In both of these schemes it is shown
that if one applies unitary operations on 2 qubits then 3 bits of
classical information can be encoded. This simply means that 8 mutually
orthogonal states can be created by applying unitary operations on
the first two qubits of a 4-qubit $W$ state. The dense coding scheme
of Pradhan \emph{et al. }\cite{pati1} is different from that of Wang
and Yan \cite{wang and yan}. This motivated us to study: How many
different ways dense coding can be done on a 4-qubit $W$ state? In
the process, we have obtained two more new sets of unitary operators,
which can be used for dense coding using 4-qubit $W$ states (see Table
\ref{tab:Scheme-1-of} and Table \ref{tab:Scheme-2-of}). Thus there
exist at least four different ways in which Alice can sent 3 bits of
classical information to Bob by sending 2 e-bits when they share a
4-qubit $W$ state. Consequently, it is straight forward to increase
the efficiency of the Yuan \emph{et al.} protocol \cite{the:high-capacity-wstate}
by replacing the set of unitary operators used by them for encoding
operations by the unitary operators used in one of the 4 allowed dense
coding schemes. The improvement in efficiency obtained here is absolutely
analogous to the increase in efficiency obtained in Cai and Li protocol
compared to the ping-pong protocol. And the modified protocol will
be essentially a DLL protocol with 4 qubit $W$ state. This can be
visualized as follows: In the modified protocol Alice prepares large
number of 4 qubit $W$ states, keeps the home photons (first two photons
of each entangled pair) with herself and sends the travel photons
to Bob. Now Alice and Bob implement a BB84 subroutine to check eavesdropping
by using half of the photons received by Bob. In absence of eavesdropping,
Alice encodes her secret message by using one of the four allowed
dense coding schemes on the remaining home photons (either she keeps
a fraction of the home photons for checking of eavesdropping or inserts
some decoy photons), reorders the sequence and sends all the home
photons to Bob. Finally, they check for eavesdropping and in absence
of eavesdropping Alice announces the exact sequence. Bob reorders
the sequence and measures the qubits in his possession in appropriate
$W$ basis to decode the secret message of Alice. This modification
would increase the efficiency as it will enable Alice to transmit
3 bits of classical information using 4-qubit $W$ state. But even
this modification would not make the $W$ state based protocols as
efficient as the 4-qubit cluster state based protocols.

At this point it is tempting to look at the general structure of the
protocols which uses rearrangement of particles and dense coding.
When we look at the general structure of these protocols then we observe
that the essential idea behind these protocols can be summarized as
follows \cite{Anindita}: In these protocols we always have a set
$Q=\{Q_{0},Q_{1},......,Q_{2^{n}-1}\}$ of $n-$partite orthonormal
state vectors which spans the $2^{n}$ dimensional Hilbert space and
a set of $m-qubit$ $m\leq n$ unitary operations $U=\{U_{0}^{i},U_{1}^{i},....U_{2^{n}-1}^{i}:U_{j}^{i}Q_{i}=Q_{j}\}$
such that the unitary operations can transform a particular element
$Q_{i}$ of set $Q$ into all the other elements of set $Q$. Now
Alice prepares multiple copies of the state vector $Q_{i}$. She encodes
$n$-bit classical message by using an encoding scheme in which $\{U_{0}^{i},U_{1}^{i},\cdots,U_{2^{n}-1}^{i}\}$
are used to encode $\{0_{1}0_{2}\cdots0_{n},\,0_{1}0_{2}\cdots1_{n},\cdots,1_{1}1_{2}\cdots1_{n}\}$
respectively. Then she rearranged the order of the particles and inserts
decoy photons (prepared in non-orthogonal states) and sends it to
Bob. If the measurement on the decoy photon does not show the existence
of Eve then Alice announces the exact sequence. Now Bob can reorder
the sequence and measure it in $\{Q_{0},Q_{1},......,Q_{2^{n}-1}\}$
basis and can unambiguously decode the message sent by Alice since
the states received by him are mutually orthogonal. Since $m=n$ is
an allowed value hence this type of protocol does not really need
dense coding. Dense coding is just a special case of the above idea.
To be precise, dense coding is possible if and only if $U_{j}^{i}$
are $m$ qubit operators, where $m<n$. In a maximal dense coding
protocol the operators are chosen in such a way that $m=\frac{n}{2}$
for even $n$ and $m=\frac{n}{2}+1$ for odd $n$. This observation
that dense coding is not required has motivated us to look into the
possibility of construction of specific DSQC and QSDC protocols using
3-qubit and 4-qubit $W$ states. In the following section such a protocol
is described and the explicit form of the unitary operators are provided.

\section{\label{sec:Generalized-protocol-of}Generalized protocol of DSQC
using symmetric $W$ states}

Here we describe a protocol which is in general valid for $n$-qubit
symmetric $W$ states but the explicit Tables of encoding operations
are provided for $n=3$ and $n=4$ only. The protocol works as follows:
\begin{description}
\item [{{Step1}}] Alice prepares a large number of copies (say $N$ copies)
of the initial state $|W_{0}\rangle$ which is a symmetric $n$-qubit
$W$ state. Then she encodes her $n$-bit classical secret message
by applying $n$-qubit unitary operators $\{U_{0},U_{1},...,U_{2^{n}-1}\}$
as described in the Table \ref{tab:Encoding-operations-for3qubitw}
for $n=3$ and in Table \ref{tab:Encoding-operations-for4qubitw}
for $n=4$. For example, to encode $0_{1}0_{2}\cdots0_{n},0_{1}0_{2}\cdots1_{n},0_{1}0_{2}\cdots1_{n-1}0_{n},\cdots\cdots\cdots,1_{1}1_{2}\cdots1_{n}$
she applies $U_{0},U_{1},U_{2},...,U_{2^{n}-1}$ respectively. The
unitary operators are chosen in such a way that the information encoded
states are mutually orthogonal. As Bob knows the initial state and
which unitary operation corresponds to what classical information,
he will be able to decode the message at the end of the protocol as
the encoded states are mutually orthogonal.
\item [{{Step2}}] Using all the $n-$partite states in her possession,
Alice creates an ordered sequence $P_{B}=[p_{1}(t_{1},t_{2},...,t_{n}),p_{2}(t_{1},t_{2},...,t_{n}),..\\ ..,p_{N}(t_{1},t_{2},...,t_{n})],$
where the subscript $1,2,...,N$ denotes the order of a $n-$partite
state $p_{i}=\{t_{1},t_{2},...,t_{n}\},$ which is in one of the $n$-partite
$W$ state $|W_{j}\rangle$ (value of $j$ depends on the encoding).
Then Alice randomly reorders the sequence $P_{B}$ of the qubits (the
actual ordering is known to Alice only) and inserts $Nn$ decoy photons%
\footnote{When $2x$ qubits (a random mix of message qubits and decoy qubits)
travel through a channel accessible to Eve and x of them are test
for eavesdropping then for any $\delta>0,$the probability of obtaining
less than $\delta n$ errors on the check qubits (decoy qubits), and
more than $(\delta+\epsilon)n$ errors on the remaining $x$ qubits
is asymptotically less than $\exp[-O(\epsilon^{2}x]$ for large x.
\cite{nielsen}. As the unconditional security obtained in quantum
cryptographic protocol relies on the fact that any attempt of Eavesdropping
can be identified. Thus to obtain an unconditional security we always
need to check half of travel qubits for eavesdropping. Thus we have
to randomly add decoy qubits whose number would be equal to the total
number of travel qubits. %
} randomly in them. Thus she creates a new sequence $P_{B}^{\prime}$,
which contains $2Nn$ photons ($Nn$ travel photons and $Nn$ decoy
photons) and sends the reordered sequence $P_{B}^{\prime}$ to Bob.
The decoy photons are prepared in a random sequence of $\{|0\rangle,|1\rangle,|+\rangle,|-\rangle\}.$
\item [{{Step3}}] After confirming that Bob has received the entire sequence,
Alice announces the position of decoy photons. Bob measures the corresponding
particles in the sequence $P_{B}^{\prime}$ by using $X$ basis or
$Z$ basis at random, here $X=\{|+\rangle,|-\rangle\}$ and $Z=\{|0\rangle,|1\rangle\}$.
After measurement, Bob publicly announces the result of his measurement
and the basis used for the measurement. Now the initial state of the
decoy photon as noted by Alice during preparation and the measurement
outcome of Bob should coincide in all such cases where Bob has used
the same basis as was used to prepare the decoy photon. Alice can
compute the error rate and check whether it exceeds the predeclared
threshold or not. If it exceeds the threshold, then Alice and Bob
abort this communication and repeat the procedure from the beginning.
Otherwise they go on to the next step. So all intercept resend attack
will be detected in this step and even if eavesdropping has happened
Eve will not obtain any meaningful information about the encoding
operation executed by Alice as the encoded information is randomized
by the rearrangement of order of the particles.
\item [{{Step4}}] Alice announces the exact sequence.
\item [{{Step5}}] Bob appropriately orders his sequence and measures
his qubits in $W$ basis. This deterministically decodes the information
sent by Alice.
\end{description}

\subsection{\label{sub:How-to-convert}How to convert this protocol into a QSDC
protocol?}

The above protocol is a protocol of DSQC as Alice needs to announce
the actual order of the sequence. Rearrangement of particle ordering
may be avoided by sending the encoded states in $n$-steps and by
checking eavesdropping after each step. Assume that Alice first sends
a sequence of all the first qubits with $N$ decoy photons, if no
eavesdropping is traced then only she sends the sequence of second
photons and so on. Then the DSQC protocol will be reduced to a QSDC
protocol as no classical information will be required for dense coding.
The previous protocol can be easily generalized to a QSDC protocol.
To do so, we just need to modify Step 2-4 in the above protocol. In
the modified protocol, after Step 1 (i.e. after the encoding is done)
Alice prepares $n$ sequences: $P_{Bi}=[p_{1}(t_{i}),p_{2}(t_{i}),....,p_{N}(t_{i})]$
with all the $i^{th}$ photons. To be precise, she prepares, $P_{B1}=[p_{1}(t_{1}),p_{2}(t_{1}),....,p_{N}(t_{1})],$
with all the first qubits, $P_{B2}=[p_{1}(t_{2}),p_{2}(t_{2}),....,p_{N}(t_{2})],$
with all the second qubits and so on. She prepares $Nn$ decoy photons
as in Step 2 of the previous protocol and inserts $N$ decoy photons
randomly into each of the $n$ sequences prepared by her. This creates
$n$ extended sequences ($P_{B1+N},\, P_{B2+N},\, P_{B3+N}$) each
of which contain $2N$ qubits. Then she sends the first sequence $P_{B1+N}$
to Bob. After confirming that Bob has received the entire sequence,
she announces the position of the decoy photons and checks eavesdropping.
If eavesdropping is found she truncates the protocol otherwise she
sends the second sequence $P_{B2+N}$ to Bob and checks for eavesdropping
and if no eavesdropping is found then she sends the third sequence
and check for eavesdropping and the process continues. Now Bob can
measure the final states in appropriate basis and obtain the message
sent by Alice. Since Eve can not obtain more than 1 qubit of a $n$-partite
state (as we are sending the qubits one by one and checking for eavesdropping
after each step) she has no information about the encoded state and
consequently this direct quantum communication protocol is secure.
Thus the rearrangement of particle order is not required if we do
the communication in multiple steps. Further, since no quantum measurement
is done at Alice's end and rearrangement of particle order is not
required, this protocol does not require any classical communication
for the decoding operation. Thus it is a QSDC protocol. Its efficiency
will be naturally higher than the previous protocol. This is so because
here Alice does not need to disclose the actual sequence and consequently
the amount of classical communication required for decoding of the
message is reduced. But this increase in qubit efficiency is associated
with a cost. This QSDC protocol will be slow as Alice has to communicate
in steps and has to check eavesdropping in the sequence before she
can send the next sequence.

\subsection{\label{sub:QKD-using-}QKD using $W$ states.}

It is obvious that a DSQC or QSDC scheme can be used for QKD. This
is so because instead of sending a meaningful message, Alice can always
decide to send a set of random bits (key). In such situation both
our DSQC and QSDC protocol will reduce to QKD protocol. It is relevant
to mention this simple idea here because Hwang \emph{et al}. \cite{hwang-hwang-Tsai}
have presented their protocol as {}``quantum key distribution protocol
using dense coding of 3-qubit W state.'' Their protocol is essentially
a DSQC protocol in which Alice sends a random key instead of a meaningful
message. Thus the improvement in efficiency of a DSQC and QSDC protocol
achieved above essentially increases the efficiency (key generation
rate) of the corresponding QKD protocol.

\section{\label{sec:Efficiency-analysis}Efficiency analysis}

In the existing literature, two analogous but different parameters
are used for analysis of efficiency of DSQC and QSDC protocols. The
first one is simply defined as \begin{equation}
\eta_{1}=\frac{c}{q}\label{eq:efficency 1}\end{equation}
 where $c$ denotes the total number of transmitted classical bits
(message bits) and $q$ denotes the total number of qubits used \cite{the:C.-W.-Tsai,hwang-hwang-Tsai}.
This simple measure does not include the classical communication that
is required for decoding of information in a DSQC protocol. Consequently
it is a weak measure. Another measure \cite{defn of qubit efficiency}
that is frequently used and which includes the classical communication
is given as \begin{equation}
\eta_{2}=\frac{c}{q+b}\label{eq:efficiency 2}\end{equation}
 where $b$ is the number of classical bits exchanged for decoding
of the message (classical communications used for checking of eavesdropping
is not counted). It is straight forward to visualize that $\eta_{1}=\eta_{2}$
for all QSDC protocols but $\eta_{1}>\eta_{2}$ for all DSQC protocols.
Now in our protocol of DSQC, as $n$-bit of classical information
is sent by $n$-qubits and equal number (i.e. $n$) of decoy qubits
so we have $c=n$ and $q=2n$. Further to disclose the actual order
we need $n$-bit of classical information. Thus $b=n.$ Therefore,
for DSQC protocol we have $\eta_{1}=\frac{1}{2}$ and $\eta_{2}=\frac{1}{3}$
and similarly for QSDC protocol we have $\eta_{1}=\eta_{2}=\frac{1}{2}.$

Now it is important to note that we can not send more than $n$-bit
of classical information by sending $n$ qubits through the channel
and to obtain an unconditional security, we always need to check half
of the travel qubits for eavesdropping. These two facts leads to a
restriction $q\geq2c$ and thus $\eta_{1}\leq\frac{1}{2}$. Further,
since in the DSQC protocol the entire sequence is disordered so to
disclose the actual order Alice needs to use $b=\frac{q}{2}$ bits
of classical information. Consequently for DSQC protocols $\eta_{2}\leq\frac{1}{3}$.
Thus the maximum efficiency of a DSQC protocol can be $33\%$ and
that of a QSDC protocol can be $50\%$ (using $\eta_{2}$ as a measure).

The above idea is used to compute the qubit efficiency $\eta_{1}$
and $\eta_{2}$ of the existing protocols of DSQC and QSDC that uses
$W$ states. The same is summarized in Table \ref{tab:Comparison-of-effiiency}.
It is interesting to note that the qubit efficiency $(\eta_{2})$
of the recently proposed 4-qubit $W$ state based protocol of Yuan
\emph{et al}. \cite{the:high-capacity-wstate} and 3-qubit based protocol
of Hwang \emph{et al}. \cite{hwang-hwang-Tsai} are $22.22\%$. Similarly
a modified version of 4-qubit $W$ state based protocol of Guo \emph{et
al.} \cite{guo} protocol%
\footnote{The original protocol is insecure as Eve can obtain substantial amount
of information before being detected. Inclusion of rearrangement of
particle ordering can make the protocol secure. But that would increase
the amount of classical communication and consequently decrease the
efficiency. %
} has an efficiency $20\%.$ Now if we use the encoding operations
described in Table \ref{tab:Scheme-1-of} or in Table \ref{tab:Scheme-2-of}
in the DLL protocol \cite{Deng protocol} then the efficiency of the
4 qubit $W$ state based DSQC protocol can be increased to $30\%$.
Now we note that the qubit efficiency of the DSQC protocol proposed
in this work is $33.33\%$ and that of its QSDC counter part is $50\%.$
Thus the proposed protocols are maximally efficient and the improvement
in the efficiency is considerable. Thus it is possible to construct
maximally efficient protocols of DSQC and QSDC using 3-qubit and 4-qubit
$W$ states, when the encoding is done by the unitary operations described
in Table \ref{tab:Encoding-operations-for3qubitw} and Table \ref{tab:Encoding-operations-for4qubitw}.

\begin{table}
\begin{centering}
\begin{tabular}{|>{\centering}p{3cm}|>{\centering}p{3in}|}
\hline
Unitary operators applied on $1^{st}$ and $2^{nd}$ qubit  & $|W\rangle_{0}=\frac{1}{2}(|0001\rangle+|0010\rangle+|0100\rangle+|1000\rangle)$\tabularnewline
\hline
X$\otimes$I  & $\frac{1}{2}(|1001\rangle+|1010\rangle+|1100\rangle+|0000\rangle)$\tabularnewline
\hline
iY$\otimes$Z  & $\frac{1}{2}(-|1001\rangle-|1010\rangle+|1100\rangle+|0000\rangle)$\tabularnewline
\hline
Z$\otimes$X  & $\frac{1}{2}(|0101\rangle+|0110\rangle+|0000\rangle-|1100\rangle)$\tabularnewline
\hline
I$\otimes$iY  & $\frac{1}{2}(-|0101\rangle-|0110\rangle+|0000\rangle-|1100\rangle)$\tabularnewline
\hline
I$\otimes$Z  & $\frac{1}{2}(|0001\rangle+|0010\rangle-|0100\rangle+|1000\rangle)$\tabularnewline
\hline
Z$\otimes$I  & $\frac{1}{2}(|0001\rangle+|0010\rangle+|0100\rangle-|1000\rangle)$\tabularnewline
\hline
iY$\otimes$iY  & $\frac{1}{2}(|1101\rangle+|1110\rangle-|1000\rangle-|0100\rangle)$\tabularnewline
\hline
X$\otimes$X  & $\frac{1}{2}(|1101\rangle+|1110\rangle+|1000\rangle+|0100\rangle)$\tabularnewline
\hline
\end{tabular}
\par\end{centering}

\caption{\label{tab:Scheme-1-of}Scheme 1 of dense coding operation on 4-qubit
$W$ state.}

\end{table}

\begin{table}
\begin{centering}
\begin{tabular}{|>{\centering}p{3cm}|>{\centering}p{3in}|}
\hline
Unitary operators applied on $1^{st}$ and $2^{nd}$ qubit  & $|W\rangle_{0}=\frac{1}{2}(|0001\rangle+|0010\rangle+|0100\rangle+|1000\rangle)$\tabularnewline
\hline
I$\otimes$X  & $\frac{1}{2}(|0101\rangle+|0110\rangle+|0000\rangle+|1100\rangle)$\tabularnewline
\hline
Z$\otimes$iY  & $\frac{1}{2}(-|0101\rangle-|0110\rangle+|0000\rangle+|1100\rangle)$\tabularnewline
\hline
iY$\otimes$I  & $\frac{1}{2}(-|1001\rangle-|1010\rangle-|1100\rangle+|0000\rangle)$\tabularnewline
\hline
X$\otimes$Z  & $\frac{1}{2}(|1001\rangle+|1010\rangle-|1100\rangle+|0000\rangle)$\tabularnewline
\hline
I$\otimes$Z  & $\frac{1}{2}(|0001\rangle+|0010\rangle-|0100\rangle+|1000\rangle)$\tabularnewline
\hline
Z$\otimes$I  & $\frac{1}{2}(|0001\rangle+|0010\rangle+|0100\rangle-|1000\rangle)$\tabularnewline
\hline
iY$\otimes$iY  & $\frac{1}{2}(|1101\rangle+|1110\rangle-|1000\rangle-|0100\rangle)$\tabularnewline
\hline
X$\otimes$X  & $\frac{1}{2}(|1101\rangle+|1110\rangle+|1000\rangle+|0100\rangle)$\tabularnewline
\hline
\end{tabular}
\par\end{centering}

\caption{\label{tab:Scheme-2-of}Scheme 2 of dense coding operation on 4-qubit
$W$ states.}

\end{table}

\begin{table}
\begin{centering}
\begin{tabular}{|c|>{\centering}p{2in}|}
\hline
Unitary operators  & 3-qubit $W$ state \tabularnewline
\hline
$U_{0}=I\otimes I\otimes I$  & $\frac{1}{\sqrt{3}}(+\left|001\right\rangle +\left|010\right\rangle +\left|100\right\rangle )$\tabularnewline
\hline
$U_{1}=X\otimes I\otimes I$  & $\frac{1}{\sqrt{3}}(+\left|101\right\rangle +\left|110\right\rangle +\left|000\right\rangle )$\tabularnewline
\hline
$U_{2}=iY\otimes X\otimes I$  & $\frac{1}{\sqrt{3}}(-\left|111\right\rangle -\left|100\right\rangle +\left|010\right\rangle )$\tabularnewline
\hline
$U_{3}=iY\otimes iY\otimes X$  & $\frac{1}{\sqrt{3}}(+\left|110\right\rangle -\left|101\right\rangle -\left|011\right\rangle )$\tabularnewline
\hline
$U_{4}=Z\otimes X\otimes iY$  & $\frac{1}{\sqrt{3}}(+\left|010\right\rangle -\left|001\right\rangle +\left|111\right\rangle )$\tabularnewline
\hline
$U_{5}=I\otimes iY\otimes I$  & $\frac{1}{\sqrt{3}}(-\left|011\right\rangle +\left|000\right\rangle -\left|110\right\rangle )$\tabularnewline
\hline
$U_{6}=X\otimes I\otimes iY$  & $\frac{1}{\sqrt{3}}(+\left|100\right\rangle -\left|111\right\rangle -\left|001\right\rangle )$\tabularnewline
\hline
$U_{7}=I\otimes Z\otimes iY$  & $\frac{1}{\sqrt{3}}(+\left|000\right\rangle +\left|011\right\rangle -\left|101\right\rangle )$\tabularnewline
\hline
\end{tabular}
\par\end{centering}

\caption{\label{tab:Encoding-operations-for3qubitw}Encoding operations for
implementation of maximally efficient DSQC and QSDC protocol using
symmetric 3-qubit $W$ state.}

\end{table}

\begin{table}
\begin{centering}
\begin{tabular}{|c|>{\centering}p{3in}|}
\hline
Unitary operators  & 4-qubit $W$ state\tabularnewline
\hline
$I\otimes I\otimes I\otimes I$  & $\frac{1}{2}(+ |0001\rangle + |0010\rangle + |0100\rangle
+ |1000\rangle)$\tabularnewline
\hline
$X\otimes I\otimes I\otimes I$  & $\frac{1}{2}(+ |0000\rangle + |1001\rangle + |1010\rangle
+ |1100\rangle)$\tabularnewline
\hline
$iY\otimes Z\otimes I\otimes I$  & $\frac{1}{2}(+ |0000\rangle - |1001\rangle - |1010\rangle
+ |1100\rangle)$\tabularnewline
\hline
$I\otimes iY\otimes I\otimes I$  & $\frac{1}{2}(+ |0000\rangle - |0101\rangle - |0110\rangle
- |1100\rangle)$\tabularnewline
\hline
$Z\otimes X\otimes I\otimes I$  & $\frac{1}{2}(+ |0000\rangle + |0101\rangle + |0110\rangle
- |1100\rangle)$\tabularnewline
\hline
$Z\otimes Z\otimes I\otimes I$  & $\frac{1}{2}(+ |0001\rangle + |0010\rangle - |0100\rangle
- |1000\rangle)$\tabularnewline
\hline
$I\otimes I\otimes X\otimes iY$  & $\frac{1}{2}(-|0001\rangle + |0010\rangle -
|0111\rangle - |1011\rangle)$\tabularnewline
\hline
$Z\otimes Z\otimes X\otimes iY$  & $\frac{1}{2}(-|0001\rangle + |0010\rangle + |0111\rangle
+ |1011\rangle)$\tabularnewline
\hline
$X\otimes iY\otimes iY\otimes iY$  & $\frac{1}{2}(+ |1101\rangle - |0111\rangle + |1011\rangle
+ |1110\rangle)$\tabularnewline
\hline
$iY\otimes X\otimes iY\otimes iY$  & $\frac{1}{2}(+ |1101\rangle + |0111\rangle - |1011\rangle
+ |1110\rangle)$\tabularnewline
\hline
$I\otimes X\otimes X\otimes iY$  & $\frac{1}{2}(+ |0110\rangle - |0101\rangle -
|0011\rangle - |1111\rangle)$\tabularnewline
\hline
$I\otimes X\otimes iY\otimes X$  & $\frac{1}{2}(- |0110\rangle + |0101\rangle - |0011\rangle
- |1111\rangle)$\tabularnewline
\hline
$iY\otimes I\otimes iY\otimes X$  & $\frac{1}{2}(+ |1010\rangle - |1001\rangle + |1111\rangle
- |0011\rangle)$\tabularnewline
\hline
$iY\otimes X\otimes I\otimes Z$  & $\frac{1}{2}(+|1101\rangle - |1110\rangle -
|1000\rangle + |0100\rangle)$\tabularnewline
\hline
$iY\otimes X\otimes Z\otimes I$  & $\frac{1}{2}(-|1101\rangle + |1110\rangle - |1000\rangle
+ |0100\rangle)$\tabularnewline
\hline
$iY\otimes I\otimes X\otimes iY$  & $\frac{1}{2}(-|0011\rangle + |1001\rangle - |1010\rangle
+ |1111\rangle)$\tabularnewline
\hline
\end{tabular}
\par\end{centering}

\caption{\label{tab:Encoding-operations-for4qubitw}Encoding operations for
implementation of maximally efficient DSQC and QSDC protocol using
symmetric 4-qubit $W$ state.}

\end{table}

\section{\label{sec:Conclusions}Conclusions}

It was well known from LM05 protocol and its variants that dense coding
is not essential for QSDC. Due to Cai and Li's protocol's success
in increasing the efficiency of ping-pong protocol by using dense
coding and because of inclusion of dense coding in all subsequent
protocols of efficient DSQC and QSDC it became a common practice to
use dense coding for designing of new protocols of direct quantum
communication. Here we have first shown that the efficiency of the
existing protocols that uses 4-qubit $W$ states can be increased
by suitable use of dense coding operations. For that purpose we have
provided two new alternative schemes for dense coding of 4-qubit $W$
states. Then we have shown that when we use a DSQC protocol based
on particle order rearrangement technique then the use of dense coding
is not essential. A simple encoding that maps the input state into
a set of mutually orthogonal states is enough. Using this fact we
have explicitly shown that $W$ state, which does not show maximal
dense coding, can be used to design maximally efficient protocol of
DSQC and QSDC. In Table \ref{tab:Encoding-operations-for3qubitw}
and Table \ref{tab:Encoding-operations-for4qubitw}, we have explicitly
provided the unitary operators required for successful implementation
of DSQC and QSDC using 3-qubit and 4-qubit $W$ states without using
dense coding. This change in strategy has considerably increased the
efficiency of protocols presented in recent past \cite{the:high-capacity-wstate,hwang-hwang-Tsai}.
To be precise, qubit efficiency is considerably improved compared
to the existing protocols of secure direct communication. This fact
can be visualized in the Table \ref{tab:Comparison-of-effiiency}.

The proposed scheme is experimentally realizable since $W$ states
can be prepared experimentally \cite{expt1} using photons and the
single qubit Pauli operators (Pauli gates), which are used to construct
multi-qubit unitary operators, can also be realized optically \cite{quantumgates}.
Further, there exist several other quantum states where maximal dense
coding is not possible (e.g. 4-qubit $Q_{4}$ and $Q_{5}$ states
\cite{pati1}). The present idea can be extended for those states
too.

\begin{table}
\begin{centering}
\begin{tabular}{|>{\centering}p{2in}|>{\centering}p{1in}|>{\centering}p{2.7cm}|c|}
\hline
Protocol  & Qubit efficiency $\eta_{1}$ in \%  & Qubit efficiency $\eta_{2}$ in \%  & Quantum states\tabularnewline
\hline
Modified DLL protocol \cite{Deng protocol}, where encoding is done
using the unitary operations described in Table \ref{tab:Scheme-1-of}
or \ref{tab:Scheme-2-of}. & 37.5 & 30 & 4-qubit $W$ state\tabularnewline
\hline
\cite{the:high-capacity-wstate}  & 33.33  & 22.22  & 4-qubit $W$ state\tabularnewline
\hline
\cite{hwang-hwang-Tsai}  & 26.67  & 22.22  & 3-qubit $W$ state\tabularnewline
\hline
Modified version of \cite{guo} & 33.33 & 20 & 4-qubit $W$ state\tabularnewline
\hline
\cite{cao and song}  & 16.67  & 14.29  & 4-qubit $W$ state\tabularnewline
\hline

Proposed DSQC protocol  & 50  & 33.33  & 3 and 4-qubit $W$\emph{ }state\tabularnewline
\hline
Proposed QSDC protocol  & 50  & 50  & 3 and 4-qubit $W$ state\tabularnewline
\hline
\end{tabular}
\par\end{centering}

\caption{\label{tab:Comparison-of-effiiency}Comparison of quantum bit efficiency
of different protocols of DSQC and QSDC that uses symmetric $W$ states.}

\end{table}

\textbf{Acknowledgment:} AP thanks Department of Science and Technology
(DST), India for support provided through the DST project No. SR/S2/LOP/2010
and the Ministry of Education of the Czech Republic for support provided
through the project CZ.1.05/2.1.00/03.0058.


\begin{thebibliography}{30}
\bibitem{bb84}C. H. Bennett and G. Brassard, in Proceedings of the
IEEE International Conference on Computers, Systems, and Signal Processing,
Bangalore, (1984) 175.

\bibitem{ekert}A. K. Ekert, Phys. Rev. Lett. \textbf{67,} (1991)
661.

\bibitem{b92}C. H. Bennett, Phys. Rev. Lett. \textbf{68,} (1992)
3121.

\bibitem{review}G. Long \emph{et al}., Front. Phys. China \textbf{2,}
(2007) 251.

\bibitem{Imoto}K. Shimizu and N. Imoto, Phys. Rev. A \textbf{60,}
(1999), 157.

\bibitem{Hillery}M. Hillery, V. Buzek and A. Bertaiume, Phys. Rev.
A \textbf{59,} (1999) 1829.

\bibitem{ping-pong}K. Bostrom and T. Felbinger, Phys. Rev. Lett.
\textbf{89,} 187902 (2002).

\bibitem{vaidman-goldenberg}L. Goldenberg and L. Vaidman, Phys. Rev.
Lett. \textbf{75}, (1995) 1239.

\bibitem{lm05}M. Lucamarini and S. Mancini, Phys. Rev. Lett. \textbf{94,}
(2005) 140501.

\bibitem{cai and li PRA}Q. Y. Cai and B. W. Li, Phys. Rev. A, \textbf{69,
}(2004) 054301.

\bibitem{Deng protocol}F. G. Deng, G. L. Long and X. S. Liu, Phys.
Rev. A \textbf{68,} (2003) 042317.

\bibitem{rearrangement of particle order}F. G. Deng and G. L. Long,
Phys. Rev. A \textbf{68,} (2003) 042315.

\bibitem{reordering1}A. D. Zhu, Y. Xia, Q. B. Fan, and S. Zhang,
Phys. Rev. A \textbf{73,} (2006) 022338.

\bibitem{the:high-capacity-wstate}H. Yuan \emph{et al.}, Int. J.
Theo. Phys. \textbf{50,} (2011) 2403.

\bibitem{the:C.-W.-Tsai}C.W. Tsai, C.R. Hsieh, and T. Hwang, Eur.
Phys. J. D \textbf{61,} (2011) 779.

\bibitem{hwang-hwang-Tsai}T. Hwang, C. C. Hwang, and C. W. Tsai,
Eur. Phys. J. D \textbf{61, }785 (2011).

\bibitem{guo}G. Zhao \emph{et al.}, Procedia Engineering \textbf{29,
}(2012) 568.

\bibitem{wang and yan}M. Y. Wang and F. L. Yan, Chin. Phys. B\textbf{
20,} 120309 (2011).

\bibitem{pati1}B. Pradhan, P. Agrawal and A. K. Pati, arXiv:0705.1917v1,
{[}quant-ph{]} (2007).

\bibitem{defn of qubit efficiency}A. Cabello, Phys. Rev. Lett. \textbf{85,}
(2000) 5635.

\bibitem{opt com-2010}C. W. Tsai and T. Hwang, Optics Comm. \textbf{283,}
(2010) 4397.

\bibitem{wang}X. W. Wang, Quant. Inf. Pro. \textbf{8,} (2009) 431.

\bibitem{cao and song}H. J. Cao, H. S. Song, Chin. Phys. Lett. \textbf{23,}
(2006) 290.

\bibitem{wang-Zhang-tang}J. Wang, Q. Zhang and C. J. Tang, Commn.
Theor. Phys. (Beijing, China)\textbf{ 48,} (2007) 637.

\bibitem{dong 1}L. Dong \emph{et al}., Commun. Theor. Phys.\textbf{
50,} (2008) 359.

\bibitem{yuan-two-step}H. Yaun \emph{et al}., Commun. Theor. Phys.
\textbf{55,} 984 (2011).

\bibitem{Anindita}A. Banerjee and A. Pathak, arXiv:1201.3763v1 {[}quant-ph{]}
(2012).

\bibitem{nielsen}M. A. Nielsen and I. L. Chuang, Quantum Computation
and Quantum Information, Cambridge University Press, New Delhi (2008)
589.

\bibitem{expt1}M. Eible \emph{et al}., Phys. Rev. Lett. \textbf{92},
(2004) 077901.

\bibitem{quantumgates}Xing-Can Yao \emph{et al.}, Phys. Rev. Lett. \textbf{105},
120402 (2010).
\end{thebibliography}
\end{document}